\documentclass[12pt]{iopart}

\usepackage{cite}

\usepackage[dvipdfmx]{graphicx}

\newcommand{\bib}{\bibitem}
\newcommand{\todayd}{\the\year/\the\month/\the\day}
\newcommand{\eq}[1]{\begin{equation} #1 \end{equation}}
\newcommand{\eqa}[2]{\begin{equation} #1 \label{#2} \end{equation}}

\newcommand{\ep}{\epsilon}
\newcommand{\la}{\left\langle}
\newcommand{\ra}{\right\rangle}

\newcommand{\lr}{\leftrightarrow}

\newcommand{\balign}[1]{\begin{eqnarray} #1 \end{eqnarray}}

\newcommand{\lb}{\label}
\newcommand{\nt}{\nonumber}

\newcommand{\eqref}[1]{(\ref{#1})}
\newcommand{\eqqref}[1]{Eq.~\eqref{#1}}
\newcommand{\ffref}[1]{Fig.~\ref{f:#1}}
\newcommand{\secref}[1]{Sec.~\ref{s:#1}}

\newcommand{\figin}[4]
{\begin{figure}[tb]
\centering
\includegraphics[width= #1]{#2}
\caption{#3}
\label{f:#4}
\end{figure}}

\def \({\left(}
\def \){\right)}

\def\rnum#1{\resizebox{0.5em}{\height}{\expandafter{\romannumeral #1}}}
\def\Rnum#1{\resizebox{0.5em}{\height}{\uppercase\expandafter{\romannumeral #1}}}

\makeatletter
  \newcommand{\subsubsubsection}{\@startsection{paragraph}{4}{\z@}%
    {1.0\Cvs \@plus.5\Cdp \@minus.2\Cdp}%
    {.1\Cvs \@plus.3\Cdp}%
    {\reset@font\sffamily\normalsize}
  }
  \makeatother
\makeatletter
\def\verbatim@font{\normalfont\fontfamily{txr}\selectfont
\let\do\do@noligs
\verbatim@nolig@list}
\makeatother

\begin{document}

\title{Measurement-feedback formalism meets information reservoirs}

\author{Naoto Shiraishi$^1$, Takumi Matsumoto$^1$, Takahiro Sagawa$^2$}
\address{$^1$ Department of Basic Science, The University of Tokyo, 3-8-1 Komaba, Meguro-ku, Tokyo, Japan}
\address{$^2$ Department of Applied Physics, The University of Tokyo, 7-3-1 Hongo, Bunkyo-ku, Tokyo, Japan}
\ead{shiraishi@noneq.c.u-tokyo.ac.jp}
\vspace{10pt}
\begin{indented}
\item \todayd
\end{indented}

\begin{abstract}
There have been two distinct formalisms of thermodynamics of information:
One is the measurement-feedback formalism, which concerns bipartite systems with measurement and feedback processes, and the other is the information reservoir formalism, which considers bit sequences as a thermodynamic fuel.
In this paper, we derive a second-law-like inequality by applying the measurement-feedback formalism to information reservoirs, which provides a stronger bound of extractable work than any other known inequality in the same setup.
In addition, we demonstrate that the Mandal-Jarzynski model, which is a prominent model of the information reservoir formalism, is equivalent to a model obtained by the contraction of a bipartite system with autonomous measurement and feedback.
Our results provide a unified view on the measurement-feedback and the information-reservoir formalisms.

\end{abstract}

\maketitle

\section{Introduction}

Evaluating the amount of work extraction from reservoirs is an important issue in thermodynamics and nonequilibrium statistical mechanics.
Here, we focus on the work extraction by using information, where two distinct formalisms exist.
One formalism concerns the measurement-feedback processes between two systems:
A celebrated example is Maxwell's demon.
In this case, the generalizations of the second law and the fluctuation theorem have been obtained by including the {\it mutual information}~\cite{Cover-Thomas} between the system and the controller (demon)~\cite{Szilard, Touchette, SU2010, SU2012, Allahverdyan, Hartich, HS, Jordan2014, SS, SIKS, Toyabe, HSP}, and it has been applied to biochemical sensing~\cite{Bo, IS, Hartich2}.
We call this avenue of research as the measurement-feedback (MF) formalism.
The other formalism concerns the work extraction from bit sequences called {\it information reservoirs}, where the Shannon entropy of a long bit sequence is consumed to extract work (see also \ffref{f:MJrevisit}(a)).
A second-law-like inequality stronger than the conventional second law has been obtained in this formalism~\cite{MJ, refri, Deffner, Boyd, BS2013, BS2014, Strasberg, Strasberg2, Merhav}.
We call this as the information-reservoir (IR) formalism.
Although both of these two formalisms address the fundamental link between information and thermodynamics, the relation between the MF and the IR formalisms has not been fully understood.
In previous researches, an important observation is that the IR formalism gives a stronger inequality than the MF formalism in some setups~\cite{BS2013, BS2014, Boyd}.

In contrast, in this paper we show that the MF formalism is also applicable to IR models, and leads to a stronger second-law-like inequality than the IR formalism.
The central idea here is the additive decomposition of the total entropy production, which was first investigated in the context of a single measurement and feedback~\cite{SU2012}, and then extended to general Markov jump processes~\cite{SS}.
This decomposition assigns the {\it partial entropy production} to each transition path, for which a generalized second law is obtained.

In addition, we show that the Mandal-Jarzynski (MJ) model, a prominent IR model, can be regarded as a contracted model of an autonomous MF model even at the level of stochastic trajectories.
In this special case, the newly obtained inequality is shown to be equivalent to the inequality with the mutual-information flow in terms of the MF formalism~\cite{Allahverdyan, Hartich, Jordan2014}.
Our result would lead to the comprehensive understanding of the link between information and thermodynamics.

\section{Partial entropy production}\lb{s:partial}

We briefly review the notion of the {partial entropy production} introduced in Ref.~\cite{SS}, which is an extension of the MF formalism.
Throughout this paper, we consider Markov jump processes with discrete states and continuous time, and focus on the stationary states in the presence of a single heat bath.
We normalize the inverse temperature $\beta$ as 1 for simplicity.
The transition from a state $w'$ to another state $w$ is written as $w' \to w$, whose transition rate is denoted by $P(w' \to w)$.
The stationary distribution $P(w)$ satisfies the stationary condition 
\eq{
0=\sum_{w'} J(w' \to w):= \sum_{w'}P(w')P(w' \to w) -P(w)P(w \to w').
}
We define the total entropy production $\sigma _{\rm tot}$ with time interval $0\leq t\leq T$ as~\cite{FTreview}
\eq{
\sigma _{\rm tot}:=-\sum_{i=1}^N Q(w_{i-1}\to w_i)+s(w(T))-s(w(0)),
}
where we denoted the number of jumps by $N$, and the $i$-th jump occurs from $w_{i-1}$ to $w_i$ at time $t_i$ ($1\leq i\leq N$). 
The state at time $t$ is written as $w(t)$.
$Q (w' \to w)=-\ln \({P(w' \to w)}/{P(w \to w')}\)$ is the heat absorption by the system from the heat bath with the transition $w' \to w$, and $s(w):=-\ln P(w)$ represents the stochastic Shannon entropy at the state $w$.
We note that the local detailed balance condition~\cite{vanKampen} is assumed.

We now define the partial entropy production $\sigma _\Omega$ for a set of transitions, written as $\Omega$, which is a subset of all possible transitions.
We first define the partial probability flux at the state $w$ as $J_{\Omega}(w):=\sum_{\{w'|(w'\to w)\in \Omega\}} J(w' \to w)$.
We then define the partial entropy production $\sigma_\Omega$ as
\eq{
\sigma_\Omega := -Q_{\Omega} + \Delta s_{\Omega}.
}
Here, we defined $Q_{\Omega}:=\sum_{i=1}^{N} Q(w_{i-1} \to w_i) \delta_\Omega (w_{i-1}\to w_i )$, where $\delta_\Omega (w'\to w)$ takes $1$ if $w'\to{w}\in\Omega$ and takes $0$ otherwise.
We also defined $\Delta s_{\Omega}:=s_{\Omega ,{\rm jump}}-\int_0^T{J_\Omega (w(t))}/{P(w(t))}dt$ with $s_{\Omega ,{\rm jump}}:=\sum_{i=1}^{N}\( s(w_{i})-s(w_{i-1})\) \delta_\Omega (w_{i-1}\to w_i)$.
It is crucial that the partial entropy production $\sigma _\Omega$ is an additive decomposition of the total entropy production 
\eq{
\sigma _\Omega +\sigma _{\Omega ^{\rm c}}=\sigma  _{\rm tot},
}
where $\Omega ^{\rm c}$ is the complement of $\Omega$.
The partial entropy production satisfies the fluctuation theorem:
\eqa{
\la e^{-\sigma _\Omega}\ra =1,
}{partial-FT}
which leads to the generalized second law for the partial entropy production:
\eqa{
\la \sigma _\Omega\ra \geq 0.
}{partial-ineq}
It has been shown that the fluctuation theorem for autonomous measurement and feedback processes is obtained as a special case of \eqqref{partial-ineq}~\cite{SS}.
We further discuss the significance of this inequality in \secref{power}.

\figin{12cm}{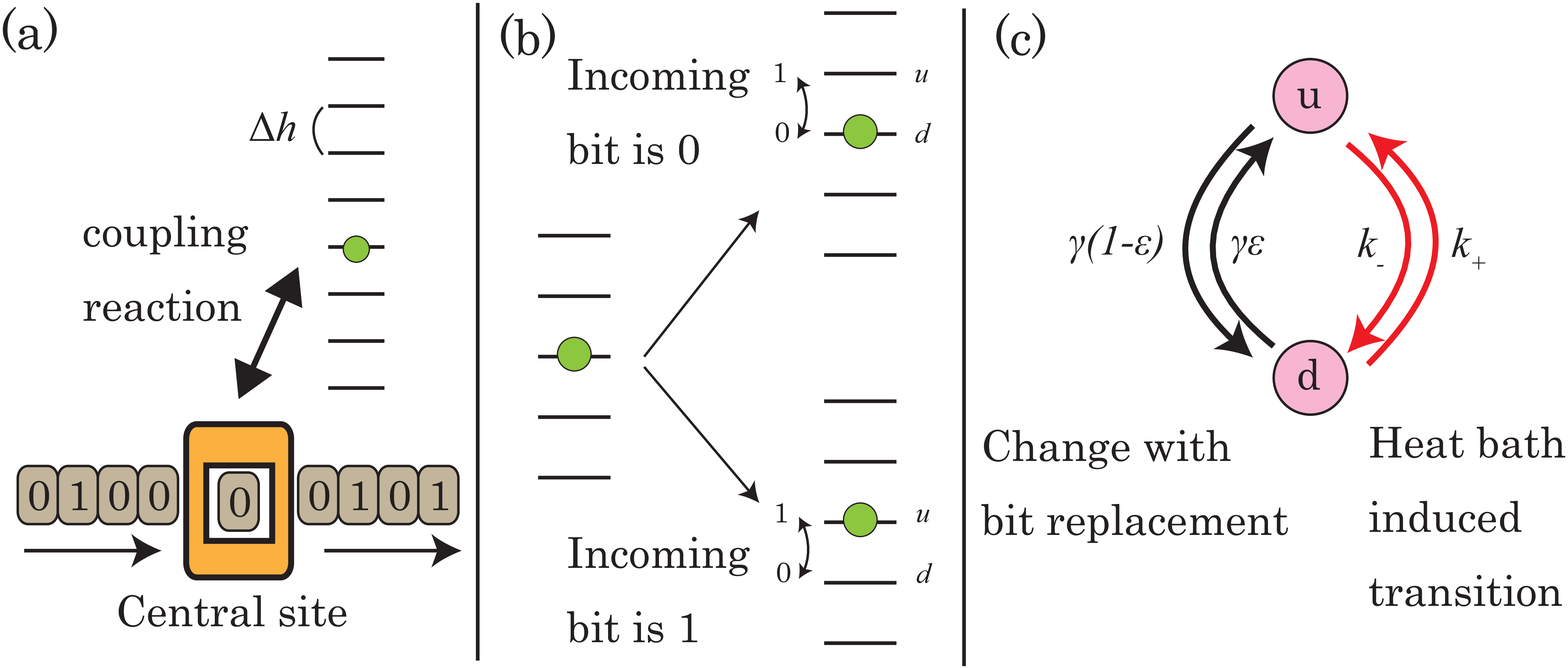}{
{(a)} Schematic of the simplified MJ model.
The horizontal lines represent the energy levels corresponding to the potential energy of the weight, and the green ball represents the height of the weight, which plays the role of a work storage.
A long bit sequence with 0 and 1 moves from left to right, which can flip their own states only at the central site (i.e., yellow cage in the figure).
The flip of a bit is coupled to the up-down of the weight.
{(b)} The labeling of the states of the weight corresponding to the incoming bit. 
If the incoming bit is 0 (1), the state of the weight is labeled as $d$ ($u$), and the higher (lower) energy level is labeled as $u$ ($d$).
{(c)} The state space and transition rates of the simplified MJ model.
The two red arrows constitute the subset $\Omega _{\rm Th}$.
}{f:MJrevisit}

\section{Mandal-Jarzynski model}\lb{s:MJ}

We here explain the MJ model which extracts work from a bit sequence~\cite{MJ, BS2013}. 
In particular, we consider a simplified version of the MJ model introduced in Ref.~\cite{BS2013}.
The simplified MJ model is a composite system of a bit sequence and a weight with the coupling reaction $(0,d)\lr (1,u)$ (see \ffref{f:MJrevisit}(a)).
Here, 0 and 1 represent states of a bit, $u$ and $d$ represent (relative) positions of the weight on ladder-like energy levels with energy interval $mg\Delta h$.
The weight plays the role of a work storage.
The bit sequence moves in one direction, from left to right, with the Poisson process with the rate $\gamma$.
The probability that the incoming bit is 1 is $\ep$.
When a new bit comes into the central site, the state of the weight is instantly labeled as $d$ ($u$) corresponding to the state of the incoming bit 0 (1), and the higher (lower) energy level is labeled as $u$ ($d$) (see \ffref{f:MJrevisit}(b)).
Under this labeling, the possible states of the composite system are only $(0,d)$ and $(1,u)$.

There are two possible transitions between $u$ and $d$ (see \ffref{f:MJrevisit}(c)).
One is the stochastic jump induced by the thermal bath with rates $P_{\rm heat}^{\rm MJ}(d\to u)=k_+$ and $P_{\rm heat}^{\rm MJ}(u\to d)=k_-$, which satisfy $k_+/k_-=e^{-mg\Delta h}$.
This is the coupling reaction of the up-down of the weight and the flip of a bit such that $(0,d)\lr (1,u)$.
We denote the set of these transitions by $\Omega _{\rm Th}$.
The other is induced by the bit stream with rates $P_{\rm bit}^{\rm MJ}(d\to u)=\gamma \ep$ and $P_{\rm bit}^{\rm MJ}(u\to d)=\gamma (1-\ep )$; from the above-mentioned rule of the state labeling, the stream of the bit sequence also changes the states $u$ and $d$.
If $\ep$ is small (i.e., the entropy of the bit sequence is low), the weight can be raised on average by increasing the entropy of the bit sequence.
The stationary probability of $u$ is given by $p_\tau :=\tau p+(1-\tau )\ep$, where $\tau :=(k_++k_-)/(k_++k_-+\gamma )$ and $p:=k_+/(k_++k_-)$.
Since $p_\tau$ is the probability that the outgoing bit is 1, the average energy gain per single incoming bit is given by $W_\tau =mg\Delta h(p_\tau -\ep )$, which we regard as work extraction.
We note that the entropy change in the work storage is irrelevant to our argument, because the fluctuation of the state of the work storage does not affect the procedure of autonomous work extraction.
This type of work storages for autonomous work extraction have been discussed in Refs.~\cite{Deffner, clock, mine}.

\section{Stronger second law from \eqref{partial-ineq}}\lb{s:power}

We now apply the fluctuation theorem with the partial entropy production, discussed in Sec.~\ref{s:partial}, to the simplified MJ model.
We set $\Omega$ in Sec.~\ref{s:partial} to $\Omega _{\rm Th}$ in Sec.~\ref{s:MJ} (the red arrows in \ffref{f:MJrevisit}(c)).
Equality \eqref{partial-FT} and inequality \eqref{partial-ineq} are then written as
\eqa{
\la e^{-\sigma _{\Omega _{\rm Th}}}\ra =1
}{MJ-FT}
and
\eqa{
((1-p_\tau)\cdot k_+ -p_\tau \cdot k_-)(-mg\Delta h-\ln p_\tau +\ln (1-p_\tau ))\geq 0.
}{MJ-ineq-mid}
By using the stationary condition $(1-p_\tau)\cdot k_+ -p_\tau \cdot k_-=p_\tau \cdot \gamma (1-\ep)-(1-p_\tau)\cdot \gamma \ep =\gamma (p_\tau -\ep)$, inequality \eqref{MJ-ineq-mid} reduces to a simple form:
\eqa{
(p_\tau -\ep )(\ln (1-p_\tau )-\ln p_\tau )\geq W_\tau .
}{restrict-MJ}
The left-hand side of \eqref{restrict-MJ} represents the entropy change induced by the heat bath.

\figin{9cm}{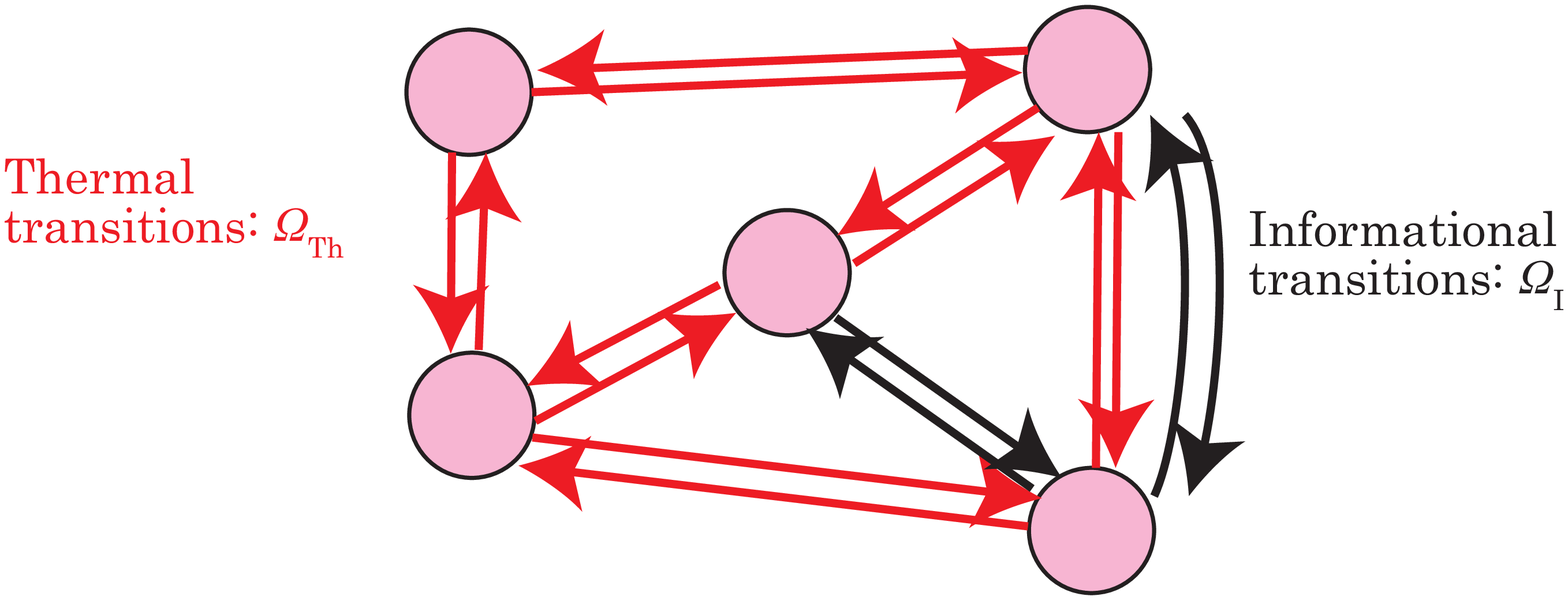}{
State space of a Markov jump process.
We divide pairs of transitions into thermal transitions (red arrows) and informational transitions (black arrows).
We consider the partial entropy production associated with the thermal transitions.
}{MJ-gen}

We next extend our foregoing argument to general Markov jump processes with states $\{ w^1,\cdots w^M\}$.
In line with Ref.~\cite{BS2014}, we divide pairs of transitions into two groups, which we refer to as {\it thermal transitions} and {\it informational transitions}.
We denote these sets of transitions by $\Omega _{\rm Th}$ and $\Omega _{\rm I}$, respectively (see \ffref{MJ-gen}).
The work is extracted during the thermal transitions.
By setting $\Omega _{\rm Th}$ to $\Omega$, \eqqref{partial-ineq} reduces to
\eqa{
Q_{\rm Th}-\sum_{w^i\lr w^j\in \Omega _{\rm I}}J(w^i\to w^j)(s(w^j)-s(w^i))\geq 0,
}{restrict-MJ-gen}
where we defined $Q_{\rm Th}:=\sum_{w^i \lr w^j \in \Omega _{\rm Th}}J(w^i\to w^j)Q(w^i\to w^j)$.

\section{Hierarchy of second-law-like inequalities}

We now show the hierarchical structure of three upper bounds on $W_\tau$ with the simplified MJ model: the conventional second law, an inequality derived from the IR formalism~\cite{MJ, BS2013}, and the newly derived inequality \eqref{restrict-MJ}.
First, we compare the conventional second law and the inequality obtained from the IR formalism.
To clarify the thermodynamic energy cost of the one-way stream of the bit sequence, we model the bit sequence by non-interacting particles with two components $0$ and $1$ (\ffref{f:tape-bath}), where the transition rates are the same as $P_{\rm bit}^{\rm MJ}$.
Particles in the bath stochastically come and go to the reaction site, which corresponds to the yellow cage of the MJ model.
The ratio of the particles $0$ and $1$ in the bath is $1-\ep :\ep$, and therefore the gain (loss) of the chemical potential with the transition $u\to d$ ($d\to u$) is $\Delta \mu =\ln ((1-\ep)/\ep )$.
Although the kinetics of the particle-bath model is equivalent to that of the bit sequence, the entropy production in the particle-bath and that in the bit sequence are different as shown below.
The conventional second law for the total system with the particle-bath is given by
\eqa{
\frac{1}{\gamma}(p_\tau \cdot P_{\rm bit}^{\rm MJ}(u\to d)-(1-p_\tau )\cdot P_{\rm bit}^{\rm MJ}(d\to u))\Delta \mu \geq W_\tau .
}{MJ-2nd}
On the other hand, the second law for the total system with the bit sequence has been derived from the IR formalism, which is given by~\cite{MJ, BS2013}
\eqa{
H(p_\tau )-H(\ep )\geq W_\tau .
}{MJ-ineq}
We refer to this inequality as the MJ inequality.
Subtracting the left-hand side (lhs) of \eqref{MJ-ineq} from the lhs of \eqref{MJ-2nd}, we obtain 
\eqa{
p_\tau \ln \frac{p_\tau}{\ep}+(1-p_\tau ) \ln \frac{1-p_\tau}{1-\ep}=:D(p_\tau ||\ep)\geq 0,
}{2nd-vs-MJ}
where $D(\cdot || \cdot )$ represents the Kullback-Leibler divergence that is non-negative~\cite{Cover-Thomas}.
Inequality \eqref{2nd-vs-MJ} confirms that the MJ inequality gives a stronger bound than the conventional second law.

Their difference $D(p_\tau ||\ep)$ characterizes an extra cost to move the bit sequence in one direction.
In the particle-bath case, the outgoing particle from the reaction site is in state $1$ with probability $p_{\tau}$, while a particle in the particle bath is in state $1$ with probability $\ep$.
Thus, when a particle leaves the reaction site to the particle bath, the entropy production in the particle bath is calculated as $-p_{\tau}\ln \ep -(1-p_{\tau})\ln (1-\ep )$.
In contrast, in the bit sequence case, the ratio of bit 1 in the bit sequences on the left side of the cage (i.e., the particle bath releasing the incoming bit) and that on the right side of the cage (i.e., the particle bath receiving the outgoing bit) are different.
Therefore, a bit on the right side of the cage is in state 1 with probability $p_{\tau}$, and when a bit leaves the reaction site to the bath, the entropy production in the bath is calculated as $-p_{\tau}\ln p_{\tau} -(1-p_{\tau})\ln (1-p_{\tau})$.
The difference between the foregoing two entropy productions is equal to $D(p_\tau ||\ep)$.
This implies that the strength of the MJ inequality comes not from the use of the bit sequence carrying information but from the one-directional operation, which inevitably needs the additional energy cost.

\figin{9cm}{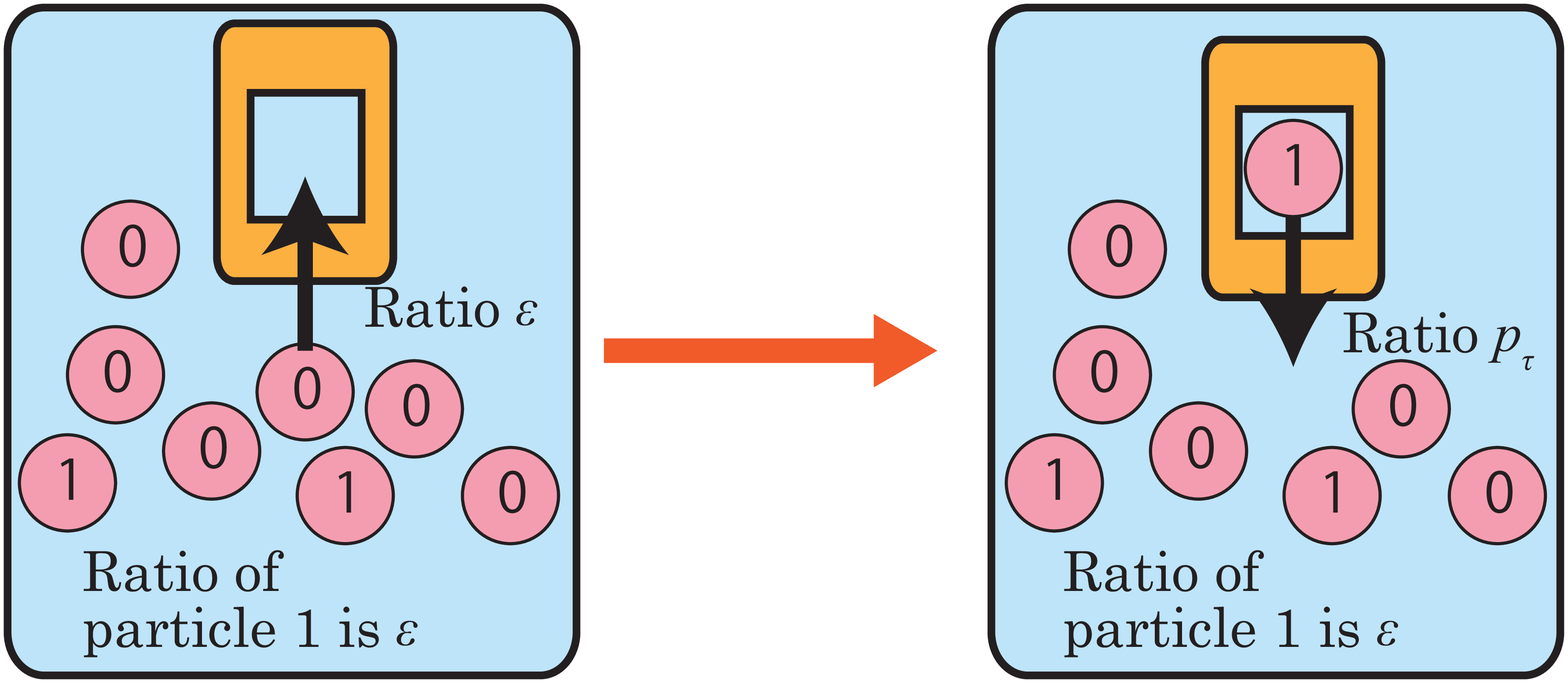}{
Schematic of the particle-bath model of a bit sequence.
A particle labeled as 0 or 1 stochastically enters the reaction site (the yellow cage) with the probability ratio $1-\ep :\ep$, and leaves the space with $1-p_{\tau}:p_{\tau}$.
}{f:tape-bath}

We now compare two inequalities \eqref{MJ-ineq} and \eqref{restrict-MJ} (i.e., the MJ inequality and the inequality derived from the partial entropy production method).
Subtracting the lhs of \eqref{restrict-MJ} from the lhs of \eqref{MJ-ineq}, we get 
\eq{
\ep \ln \frac{\ep}{p_\tau}+(1-\ep )\ln \frac{1-\ep}{1-p_\tau}=:D(\ep ||p_\tau ) \geq 0.
}
Here, $D(\ep ||p_\tau )$ characterizes the dissipation corresponding to the fact that the time evolution is not quasi-static~\cite{KPB}.
The left-hand side of the MJ inequality \eqref{MJ-ineq} is just characterized by the entropy change between the initial state (i.e., the state of the incoming bit) and the final state (i.e., the state of the outgoing bit).
Therefore, the MJ inequality does not take into account the dissipation due to the free relaxation of a bit, while inequality \eqref{restrict-MJ} can take it into account.
Such dissipation is unavoidable, as autonomous information engines inevitably involve relaxation processes, because of the very fact that they are autonomous.
This is the physical reason why the partial entropy production method gives a stronger bound of the work extraction.

More generally, we can show that inequality \eqref{restrict-MJ-gen} is a stronger inequality than the generalized version of the MJ inequality for the general IR formalism~\cite{BS2014}.
The setup for the generalized MJ inequality is the same as that for inequality \eqref{restrict-MJ-gen}.
The general version of the MJ inequality \eqref{MJ-ineq} is written as~\cite{BS2014}
\eqa{
Q_{\rm Th}+\sum_{w^i\lr w^j\in \Omega _{\rm I}}(P(w^i)+P(w^j))\gamma_{ij}(H(p_{ij})-H(\ep_{ij}))\geq 0,
}{MJ-gen}
where we defined $p_{ij}:=P(w^i)/(P(w^i)+P(w^j))$ and $\gamma _{ij}:=P(w^i\to w^j)+P(w^j\to w^i)$.
We then compare two inequalities \eqref{MJ-gen} and \eqref{restrict-MJ-gen}.
Subtracting the lhs of \eqref{restrict-MJ-gen} from lhs of \eqref{MJ-gen}, we obtain 
\eq{
\sum_{w^i\lr w^j\in \Omega _{\rm I}}(P(w^i)+P(w^j))\gamma_{ij}D(\ep_{ij}||p_{ij})\geq 0,
}
where we defined a positive constant $\ep _{ij}$ as $\ep _{ij}:=P(w^j\to w^i)/(P(w^i\to w^j)+P(w^j\to w^i))$ for each informational transition $w^i\lr w^j$.
Therefore, inequality \eqref{restrict-MJ-gen} is stronger than \eqref{MJ-gen}.

\section{Contraction from a MF model to the MJ model}

We next consider an autonomous measurement-feedback (AMF) model~\cite{HSP}, and show the equivalence of the simplified MJ model and the AMF model.
This equivalence allows us to directly compare the MJ inequality and the second-law-like inequality with the mutual information flow.
The AMF model consists of a fluctuating particle on the tilted one-dimensional discrete lattice and the delta-function-like wall (see \ffref{info-rat}(a)).
The position on the lattice is labeled as $x\in \{ 0,1\}$ in mod 2, and the delta-function-like walls are at the right side of 0 or 1, which we label as $y=e$ or $o$, respectively.
The energy difference of the particle between a site and its right neighbor is $mg\Delta h$.
The transition rate of the position of the particle due to the thermal diffusion is given by
\eq{
\cases{
P(0\to 1;o)=P(1\to 0;e)=k_+, \\
P(1\to 0;o)=P(0\to 1;e)=k_-,}
}
with $k_+/k_-=e^{-mg\Delta h}$.
The transition rate of the positions of the walls is given by 
\eq{
\cases{
P(e\to o;0)=P(o\to e;1)=\gamma \cdot \frac{e^{\Delta \mu}}{1+e^{\Delta \mu}}, \\
P(o\to e;0)=P(e\to o;1)=\gamma \cdot \frac{1}{1+e^{\Delta \mu}},}
}
which are equal to $P_{\rm bit}^{\rm MJ}(u\to d)$ and $P_{\rm bit}^{\rm MJ}(d\to u)$, respectively.
We note that $\Delta \mu$ is regarded as the chemical potential difference that drives the switching of the walls.
Owing to the chemical driving, the particle climes up against the potential bias when $mg\Delta h<\Delta \mu$.

\figin{9cm}{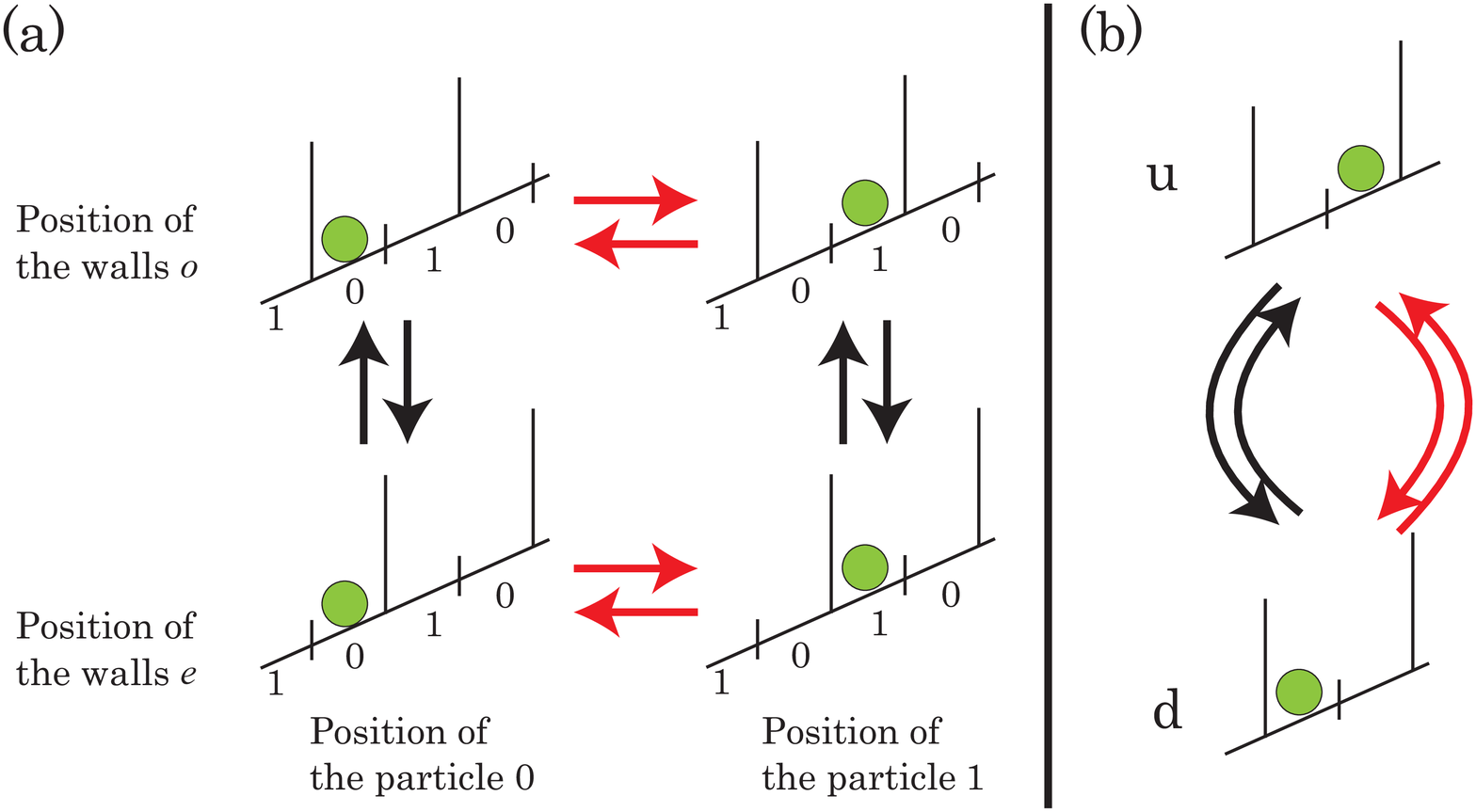}{
(a) State space of the AMF model, which consists of the tilted potential and the delta-function-like walls.
The position of the particle is labeled as 0 or 1, and the position of the walls are labeled as $e$ or $o$.
The switching of the position of the walls is driven by the chemical potential, and the position of the particle changes due to the thermal diffusion.
(b) Contraction of the AMF model.
The states $(0,o)$ and $(1,e)$ are contracted to $d$, and $(0,e)$ and $(1,o)$ to $u$.
The contracted state space and transition rates are exactly equal to those of the MJ model illustrated in \ffref{f:MJrevisit}(c).
}{info-rat}

We stress that the particle and the walls do not directly change the energy, but exchange only information.
In this sense, the AMF model is a typical example of an autonomous information-driven engine~\cite{Sekimoto, Esposito1, SS, SIKS}.
Analogous to Maxwell's demon, the walls measure the position of the particle, and change their own position depending on the measurement outcome, which is the feedback control.
The AMF model is a bipartite system with two variables $(x,y)$, and thus we can define the entropy production associated only with the particle as $\sigma _x:=-Q_x+\Delta s_x$, where $Q_x$ is the heat absorbed by the subsystem $x$, and $\Delta s_x:=-\ln P(x(T))+\ln P(x(0))$ is the change in the stochastic Shannon entropy of $x$.
Due to the presence of the autonomous measurement and feedback, $\sigma _x$ can be negative on average and does not satisfy the fluctuation theorem in the conventional form.
However, by taking into account the flow of stochastic mutual information $I(x;y)=\ln P(x,y)/P(x)P(y)$~\cite{SU2012}, we obtain the modified fluctuation theorem in the stationary state~\cite{SS}:
\eqa{
\la e^{-\sigma _x+\Delta I_x}\ra =1.
}{infoFT}
Here we defined the mutual information flow by $\Delta I_x:= I_{x,{\rm jump}}+\int_0^T{J_x(x,y)}/{P(x,y)}dt$, where $I_{x,{\rm jump}}:=\sum_{i=1}^N\( I(x_i;y_i)-I(x_{i-1};y_{i-1})\) \delta_{y_i,y_{i-1}}$ with $w_i=:(x_i, y_i)$ and Kronecker delta $\delta$, and $J_x(x,y):=\sum_{x'}J((x',y)\to (x,y))$.
We stress that \eqqref{infoFT} is a special case of \eqqref{partial-FT}:
If we apply \eqqref{partial-FT} to the AMF model by setting all transitions of $x$ to $\Omega$, we straightforwardly obtain \eqqref{infoFT}.
By applying Jensen's inequality to \eqqref{infoFT}, the generalized second law with the mutual information flow is obtained as~\cite{Allahverdyan, Hartich, Jordan2014, SS}
\eqa{
\la \dot{\sigma}_x \ra-\sum_{x,x',y}(I(x;y)-I(x';y))J((x',y)\to (x,y))\geq 0,
}{infoineq}
where $\la \dot{\sigma}_x\ra$ represents the stationary entropy production rate of $x$.
This inequality is stronger than the conventional second law of the composite system.

We now consider the contraction of the AMF model by contracting two states $(0,o)$ and $(1,e)$ to a single state $d$, and $(0,e)$ and $(1,o)$ to $u$ (see \ffref{info-rat}(b)).
This contraction means that the absolute positions of the particle and walls are replaced by the relative positions.
Since the AMF model has the two types of transition paths corresponding to the thermal diffusion of the particle and the chemical switching of the walls, its contracted model also has two types of transition paths between $u$ and $d$, which we refer to as the {thermal transition} and the {informational transition}.
Then, by denoting the transition rates of the thermal and informational transitions by $P_{\rm heat}^{\rm cAMF}$ and $P_{\rm bit}^{\rm cAMF}$ respectively, the contracted AMF model is regarded equivalent to the simplified MJ model introduced in \secref{MJ}.
Correspondingly, each stochastic trajectory in the AMF model is contracted to that in the simplified MJ model, where the transition rates are kept unchanged:
\balign{
P_{\rm heat}^{\rm MJ}(u\to d)&=P_{\rm heat}^{\rm cAMF}(u\to d), \\
P_{\rm heat}^{\rm MJ}(d\to u)&=P_{\rm heat}^{\rm cAMF}(d\to u), \\
P_{\rm bit}^{\rm MJ}(u\to d)&=P_{\rm bit}^{\rm cAMF}(u\to d), \\
P_{\rm bit}^{\rm MJ}(d\to u)&=P_{\rm bit}^{\rm cAMF}(d\to u).
}
We denote as $P^{\rm cAMF}(\Gamma )$ and $P^{\rm MJ}(\Gamma )$ the probabilities that trajectory $\Gamma$ from time 0 to $T$ and its contraction are realized in the AMF model and the simplified MJ model, respectively.
We then obtain
\eq{
P^{\rm cAMF}(\Gamma )=\frac{P^{\rm MJ}(\Gamma )}{2},
}
where the factor $1/2$ comes from the contraction of the initial state:
If the initial state after the contraction is $d$ ($u$), the initial state before the contraction is either $(0,o)$ or $(1,e)$ ($(0,e)$ or $(1,o)$) with probability $1/2$.

We next show that \eqqref{infoFT} directly reduces to \eqqref{MJ-FT}, and correspondingly, inequality \eqref{infoineq} reduces to inequality \eqref{restrict-MJ}.
Therefore, for the case of the simplified MJ model we can see the direct application of the MF formalism to an IR model without invoking the partial entropy production method.
We first note that
\balign{
&\ln P(x')-\ln P(x)+\( \ln \frac{P(x',y')}{P(x')P(y')}-\ln \frac{P(x,y)}{P(x)P(y)}\) \delta _{y,y'} \nt \\
=&\( \ln P(x')-\ln P(x)+\ln \frac{P(x',y')}{P(x')P(y')}-\ln \frac{P(x,y)}{P(x)P(y)}\) \delta _{y,y'} \nt \\
=&(\ln P(x',y')-\ln P(x,y))\delta _{y,y'}
}
holds for $x=x',y\neq y'$ or $x\neq x',y=y'$, where every line is just zero if $x=x'$, $y\neq y'$.
We then obtain
\eqa{
-\Delta s_x+I_{x, {\rm jump}}=-s_{\Omega _{\rm Th},{\rm jump}}
}{s-equiv}
for any trajectory $\Gamma$ of the AMF model.
Equality \eqref{s-equiv} implies that the two fluctuation theorems, Eqs.~\eqref{infoFT} and \eqref{MJ-FT}, are equivalent, and thus inequalities \eqref{infoineq} and \eqref{restrict-MJ} are also equivalent. 
This clarifies the physical meaning of inequality \eqref{restrict-MJ} from the viewpoint of the MF formalism:
The bit sequence plays the equivalent role to the chemical fuel that drives the autonomous measurement and feedback.

\section{Concluding remarks}
We have established the fundamental relation between the MF and the IR formalisms.
Inequality \eqref{restrict-MJ} derived from the partial entropy production method provides a stronger inequality than that derived from the IR formalism \eqref{MJ-ineq}.
Moreover, we have demonstrated that the contraction of the AMF model, to which the fluctuation theorem with the mutual information flow is applicable, is equivalent to the simplified MJ model.
This allows us the direct comparison of the inequality with the mutual information flow and the MJ inequality.

We have also shown the hierarchy of the three second-law-like inequalities for the simplified MJ model: the conventional second law \eqref{MJ-2nd}, the MJ inequality \eqref{MJ-ineq}, and the inequality \eqref{restrict-MJ} with the partial entropy production.
An important insight is that the more source of dissipation is taken into account, the stronger inequality we obtain.
From this perspective, the reason why inequality \eqref{restrict-MJ} gives the strongest bound is the following: 
It specifies the entropy production only associated with the thermal transitions, and it takes into account the dissipation induced by the free relaxation of the bit.
In general, the partial entropy production method would be useful to investigate stronger second-law-like inequalities for a variety of nonequilibrium dynamics.

In addition, we discuss a possible extension of our result to quantum cases.
Recently, the refinement of quantum thermodynamics has been intensively studied from several perspectives~\cite{clock, Campisi, Goldstein, Brandao, Esposito}.
These studies have revealed novel and fundamental aspects of the second law of thermodynamics.
In addition, both of the MF formalism~\cite{SU2008, Funo} and the IR formalism~\cite{Chapman} for quantum systems has already been discussed.
Unifying these two formalisms in light of quantum thermodynamics would be an interesting future problem.

\ack

The authors thank Sosuke Ito and Kyogo Kawaguchi for fruitful discussions and helpful comments.
NS is supported by Grant-in-Aid for JSPS Fellows Number 26-7602.
TS is supported by JSPS KAKENHI Grant No. 25800217 and No. 22340114, by KAKENHI No. 25103003 ``Fluctuation \verb#&# Structure", and by Platform for Dynamic Approaches to Living System from MEXT, Japan.


\end{document}